# The Vacuum Fluctuation Theorem:

# Exact Schrödinger Equation

# via Nonequilibrium Thermodynamics

Gerhard Grössing,

*Austrian Institute for Nonlinear Studies,*

Akademiehof,

Friedrichstrasse 10, A-1010 Vienna, Austria

e-mail: ains@chello.at

**Abstract:** Assuming that a particle of energy $\hbar\omega$ is actually a dissipative system maintained in a nonequilibrium steady state by a constant throughput of energy (heat flow), one obtains the shortest derivation of the Schrödinger equation from (modern) classical physics in the literature, and the only exact one, too.

## 1. Introduction

Ever since Albert Einstein in one of his famous papers of 1905 [1] postulated the corresponding formula, evidence has accumulated, and is nowadays a firm basis of quantum theory, that to each particle of nature one associates an energy

$$E = \hbar\omega, \qquad (1.1)$$

where $\hbar = h/2\pi$, with Planck's quantum of action $h$, and $\omega$ a characteristic angular frequency.

Surprisingly, however, this universal feature *per se* is somehow taken for granted, with not much, if any, questioning of how these oscillations, as represented by $\omega$, come about. Not even in causal, or realistic, interpretations of quantum theory, is this feature much discussed, but rather comes along only as empirical "input" into the formalism, just as in the more orthodox approaches. One could thus get the impression that the fact that particles' energies are essentially frequencies must be considered to be some kind of "axiom", i.e., an unexplainable basic feature with no prospect for a deeper understanding of its causes. However, this impression can be misleading, and, in fact, shall be dismissed here in favour of an approach that tries to present a more encompassing framework, within which said universal feature can be understood.

We are actually only confronted with these two basic options: either the quantum oscillations as mentioned above are introduced via some "axiom", or they are conceived as the results of known physical laws. In this paper we adopt the second option. It is well known that oscillations in general are the result of dissipative processes, so that the mentioned frequencies $\omega$ can be understood within the framework of nonequilibrium thermodynamics, or, more precisely, as properties of off-equilibrium steady-state systems maintained by a permanent throughput of energy.

So, we shall deal here with a "hidden" thermodynamics, out of which the known features of quantum theory should *emerge*. (This says, among other things, that we do *not* occupy ourselves here with the usual quantum versions of thermodynamics, out of which classical thermodynamics is assumed to emerge, since we intend to deal with a level "below" that of quantum theory, to begin with.)

Of course, there is *a priori* no guarantee that nonequilibrium thermodynamics is in fact operative on the level of a hypothetical sub-quantum "medium", but, as will be shown here, the straightforwardness and simplicity of how the exact central features of quantum theory emerge from this ansatz will speak for themselves. Moreover, one can even reverse the doubter's questions and ask for compelling reasons, once one does assume the existence of some sub-quantum domain with real physics going on in it, why this medium should *not* obey the known laws of, say, statistical mechanics. For, one also has to bear in mind, a number of physical systems exhibit very similar, if not identical, behaviours at vastly different length scales. For example, the laws of hydrodynamics are successfully applied even to the largest structures in the known universe, as well as on scales of kilometres, or centimetres, or even in the collective behaviour of quantum systems. In short, although there is no *a priori* guarantee of success, there is also no principle that could prevent us from applying present-day thermodynamics to the sub-quantum regime.

In fact, this is the program of the present paper: We, too, take equation (1.1) as the (only) empirical input to our approach, but we also try to understand how this can come about. For this, we study nonequilibrium thermodynamics. That is all we need in order to arrive at quite astounding results. In other words, what is proposed here can be considered also as a gedanken experiment: what if our knowledge of classical physics (including wave mechanics and nonequilibrium thermodynamics) of today had been available 100 years ago? The answer is as follows: One could have thus, without *any* further assumptions or any *ad hoc* choices of constants, derived the exact Schrödinger equation, both for conservative and nonconservative systems, using only *universal properties* of oscillators and nonequilibrium thermostatting. It is particularly the latter feature which is rather appealing, since the use of universality properties guarantees *model independence*. That is, it will turn out unnecessary to have much knowledge about the detailed sub-quantum mechanisms, as the universal properties of the systems in question will be shown to suffice to obtain the results looked for. Moreover, the approach to be presented here not only re-produces the Schrödinger equation, but also puts forward some new results, such as the sub-quantum fluctuation theorem, which can thus help shed light on problems not properly understood today within the known quantum formalism.

The paper is structured as follows. In Chapter 2, a short review is given of some results from nonequilibrium thermodynamics, which are particularly useful for our purposes. Chapter 3 then presents the application of the corresponding sub-quantum modelling of conservative systems, thus providing a straightforward derivation of the Schrödinger equation from modern classical physics. It is claimed that this represents the shortest derivation of the Schrödinger equation from classical physics in the literature, and the only exact one, too. In Chapter 4, then, the scheme is extended to include the Schrödinger equation for integrable nonconservative systems. Finally, the more encompassing scope of the present approach is presented, culminating in a formulation and discussion of the "vacuum fluctuation theorem", with particular emphasis being put on possible applications for a better understanding of quantum mechanical nonlocality.

## 2. Some Results from Nonequilibrium Thermodynamics

In the thermodynamics of small objects, the interactions with their environments are dominated by thermal fluctuations. Since the 1980ies, new experimental and theoretical tools have been developed to provide a firm basis for a theory of the nonequilibrium thermodynamics of small systems. Most characteristic for such systems are the irreversible heat losses between the system and its environment, the latter typically being a thermal bath. In recent years, a unified treatment of arbitrarily large fluctuations in small systems has been achieved by the formulation of so-called fluctuation theorems (FT). One type of FT has been developed by G. Gallavotti and E. Cohen [2] and deals with steady-state systems.

Steady-state systems are characterized by an external agent continuously producing heat which thus contributes to the small system's heat bath. The rate at which the system exchanges heat with this bath is given by the entropy production $\sigma = \Delta S_e / t$, where the entropy $S_e = \Delta Q / T$, with $T$ being the temperature and $t$ the time interval over which the system exchanges the heat $\Delta Q$. Gallavotti and Cohen associate the entropy production with a time-dependent probability distribution in phase space, $P_t(\sigma)$, and their FT provides an expression for the ratio of the probability of absorbing a given amount of heat versus that of releasing it:

$$\lim_{t\to\infty} \frac{k}{t} \ln\left(\frac{P_t(\sigma)}{P_t(-\sigma)}\right) = \sigma, \qquad (2.1)$$

where $k$ is Boltzmann's constant.

Practically, Eq. (2.1) also holds to good approximation for finite times, i.e., as long as $t$ is much greater than a given decorrelation time. Eq. (2.1) expresses the fact that nonequilibrium steady state systems *on average* always tend to dissipate heat rather than absorb it. Nevertheless, it also gives an exact probability for heat absorption (negative $\sigma$), which still is non-zero. (For an excellent review, see Evans and Searles [3].)

Related to Eq. (2.1), but actually more apt for our purposes, is a FT given by Williams, Searles, and Evans in 2004. [4] They consider what happens to a nonequilibrium dissipative system, where the initial conditions are assumed to be known, and where the system is maintained at a constant temperature. (We recall that, as an application, we want to treat the particles of quantum mechanics as such "small systems", and it is natural to start with the suggestion that they are held at some constant temperature, at least in the free-particle case.)

If this small system is surrounded by a heat bath, and if the heat capacity of this thermal reservoir is much greater than that of the system, one can "expect the system to relax to a nonequilibrium quasi-steady-state in which the rate of temperature rise for the … system is so small that it can be regarded as being zero." [4] In their paper, Williams *et al.* give a detailed analysis to show how their "transient" fluctuation theorem (TFT) is independent of the precise mathematical details of the thermostatting mechanism for an infinite class of fictitious time reversible deterministic thermostats. They thus prove the factual independence of their TFT from the thermostatting details, a fact which we denote as "universality of thermostatting" for nonequilibrium steady-state systems.

The kinetic temperature of the heat reservoir is defined by

$$kT = \frac{1}{DmN_r} \sum_{i=1}^{N_r} \mathbf{p_i} \cdot \mathbf{p_i} , \qquad (2.2)$$

where $D$ is the Cartesian dimension of the system, $N_r$ the number of reservoir particles, $\mathbf{p_i}$ their momenta and $m$ their individual masses. Since the reservoir is very large compared to the small dissipative system, one can safely assume that the momentum distribution in this region is given by the usual Maxwell-Boltzmann distribution. This corresponds to a "thermostatic" regulation of the reservoir's temperature. Now, if the phase space distribution function of trajectories $\mathbf{\Gamma}(t)$, i.e., $f(\mathbf{\Gamma}(t))$, for the thermostatted system is known, Williams *et al.* show how the TFT can be applied. Instead of using the entropy production $\sigma$ as in Eq. (2.1), the TFT now has to be formulated with the aid of a more generalized version of it, the so-called dissipation function $\overline{\Omega_t}$. It is defined by the following equation [4]:

$$\overline{\Omega_t}t := \int_0^t ds\, \Omega(\mathbf{\Gamma}(s)) \equiv \ln \frac{f(\mathbf{\Gamma}(0),0)}{f(\mathbf{\Gamma}(t),0)} - \int_0^t \Lambda(\mathbf{\Gamma}(s))\, ds , \qquad (2.3)$$

where $f(\mathbf{\Gamma}(0),0)$ is the initial $(t=0)$ distribution of the particle trajectories $\mathbf{\Gamma}$, $\mathbf{\Gamma}(t)$ is the corresponding state at time $t$, $f(\mathbf{\Gamma}(t),0)$ the initial distribution of those time evolved states, and $\Lambda(\mathbf{\Gamma}) \equiv \partial \dot{\mathbf{\Gamma}}/\mathbf{\Gamma}$ the phase space compression factor. Similar to Eq. (2.1), the TFT now provides the probability ratio

$$\frac{p(\overline{\Omega_t} = A)}{p(\overline{\Omega_t} = -A)} = e^{At} . \qquad (2.4)$$

The notation $p(\overline{\Omega_t} = A)$ is used to denote the probability that the value of $\overline{\Omega_t}$ lies in the range from $A$ to $A + dA$, and $p(\overline{\Omega_t} = -A)$ refers to the range from $-A$ to $-A - dA$. Because of the equilibrium distribution of the thermostat, or, equivalently, because the energy lost to the thermostat can be regarded as heat, the phase space compression factor is essentially given by the heat transfer $\Delta Q$,

$$\int_0^t \Lambda(\mathbf{\Gamma}(s)) \, ds \equiv \frac{\Delta Q}{kT} , \qquad (2.5)$$

and the first expression on the r.h.s. of Eq. (2.3) is equal to the change of the total energy $\Delta H$, i.e.,

$$\ln \frac{f(\mathbf{\Gamma}(0),0)}{f(\mathbf{\Gamma}(t),0)} = \frac{\Delta H}{kT} . \qquad (2.6)$$

The authors are able to show that generally, when the number of degrees of freedom in the reservoir is much larger than the number of degrees of freedom in the small system of interest, the dissipation function is equal to the work $\Delta W$ applied to the system,

$$\overline{\Omega_t} t = \frac{1}{kT}(\Delta H(t) - \Delta Q(t)) = \frac{\Delta W(t)}{kT} . \qquad (2.7)$$

By definition, the latter is given by [4]

$$\Delta W = -\int_0^t ds \, \mathbf{J}(\mathbf{\Gamma}(s)) \, \tilde{V} \cdot \mathbf{F}_e , \qquad (2.8)$$

where the dissipative field $\mathbf{F_e}$ does work on the system by driving it away from equilibrium, $\mathbf{J}$ is the so-called dissipative flux, and $\widetilde{V}$ the volume of interest. This work is converted into heat, which is in turn removed by the thermostatted reservoir particles, thus maintaining a nonequilibrium steady state.

Finally, substituting Eq. (2.8) into Eq. (2.4) provides the TFT implied by universal thermostatting (with the bars denoting averaging) [4]:

$$\frac{p\left(-\frac{1}{kT}\overline{\mathbf{J_t}\cdot\mathbf{F_e}}=A\right)}{p\left(-\frac{1}{kT}\overline{\mathbf{J_t}\cdot\mathbf{F_e}}=-A\right)}=e^{A\widetilde{V}t} \quad . \tag{2.9}$$

## 3. Merging thermodynamics with wave mechanics: Emergence of quantum behaviour

### 3.1 The basic assumptions

From the beginning, early in the twentieth century, and onwards, quantum phenomena have been characterized by both particle and wave aspects. Let us accept Eq. (1.1), $E=\hbar\omega$, as the main "empirical input" to our approach, and note as an aside that the oscillations indicated by the frequency $\omega$ can be considered as those of a carrier wave, which, depending on an observer's rest frame, are modulated such that the free particle's velocity is given by the group velocity of the associated wave. (The "free particle" is an idealization, with the particle considered to be un-affected by the thermodynamic "disturbances", which will be introduced below. Still, in many cases, the *average* particle velocity will equal the group velocity even after those disturbances are accounted for.) From classical wave mechanics we then know that $v = \frac{d\omega}{dk}$, but we generally also have that $v = \frac{dE}{dp}$ (i.e., with wave number $k$ and momentum $p$), so that by comparison we thus obtain with Eq. (1.1) that the particle's momentum is given by de Broglie's relation $\mathbf{p} = \hbar \mathbf{k}$.

So, one can imagine a particle as an oscillating entity which is in contact with its surroundings via a wave-like dynamics related to its frequency $\omega$. As we want to consider a classical wave, we can note that the probability density $P(\mathbf{x},t)$ for the presence of such a particle (which thus is equal to the detection probability density) is such that it coincides with the wave's intensity $I(\mathbf{x},t) = R^2(\mathbf{x},t)$, with $R(\mathbf{x},t)$ being the wave's (real-valued) amplitude (**Assumption 1**):

$$P(\mathbf{x},t) = R^2(\mathbf{x},t), \text{ with normalization } \int P \, d^n x = 1 . \qquad (3.1.1)$$

Now let us propose the central argument of our approach. We assume that a sub-quantum (nonequilibrium) thermodynamics provides the correct statistical mechanics responsible for the understanding of the oscillatory behaviour of a single particle on the quantum level. The "language" used is of course one of ensembles of (sub-quantum) particles, and the task is to find the appropriate transition to the ensemble behaviour of many particles (e.g., one particle in many consecutive runs of an experiment) on the quantum level. We propose that by merging the sub-quantum thermodynamics with classical wave mechanics, the emergence of quantum behaviour can be exactly modelled.

To do so, we must ask how the probability densities of a particle on the quantum level are constructed from the sub-quantum distribution functions (i.e., of $N$–particle statistical mechanics). We propose that the temporal evolution of the quantum particle's probability density in configuration space is an emerging property of the system's description based on the underlying temporal evolution of the corresponding sub-quantum distribution function, i.e.,

$$P(\mathbf{x},t) = \frac{f(\mathbf{\Gamma}(t),0)}{f(\mathbf{\Gamma}(0),0)} P(\mathbf{x},0) .$$

The equilibrium distribution $f(\mathbf{\Gamma}(t),0) = f(\mathbf{\Gamma}(0),0) e^{-\frac{\Delta H}{kT}}$ according to Eq. (2.6) is therefore assumed to be reflected also in the distribution $P(\mathbf{x},t) = P(\mathbf{x},0) e^{-\frac{\Delta H}{kT}}$.

In other words, the second "input" to our theory, is provided by the following *proposition of emergence* (**Assumption 2**):

the relation between the distribution functions referring to the trajectories at the times $0$ and $t$, respectively, on the sub-quantum level is mirrored by the corresponding relation between the probability densities on the quantum level:

$$\frac{f(\mathbf{\Gamma}(t),0)}{f(\mathbf{\Gamma}(0),0)} = \frac{P(\mathbf{x},t)}{P(\mathbf{x},0)} . \tag{3.1.2}$$

In Eq. (3.1.2) it is proposed that the many microscopic degrees of freedom associated with the subquantum medium are recast into the more "macroscopic" properties that characterize a collective wave-like behaviour on the quantum level. (This will imply that the buffeting effects of the surroundings on the particle are represented by a fluctuating force, as we shall see below.) Similar to the thermodynamics of a colloidal particle in an optical trap [5], the relevant description of the system is no longer given by the totality of all coordinates and momenta of the microscopic entities, but is reduced to only the particle coordinates.

This "emergence" of the ratio (3.1.2) on the quantum level can be justified on dynamical grounds. Assuming that the probability density (3.1.1) obeys the usual continuity equation, i.e.,

$$\frac{\partial}{\partial t}P + \nabla \cdot (\mathbf{v}P) = 0, \quad (3.1.3)$$

with solutions

$$P(t) = P(0)e^{-\int_0^t (\nabla \cdot \mathbf{v})dt}, \quad (3.1.4)$$

we see that the exponent in Eq. (3.1.4) exactly matches a familiar form of the phase space compression factor, i.e.,

$$\Lambda(\mathbf{x},t) = \nabla \cdot \mathbf{v}. \quad (3.1.5)$$

As in this chapter we assume, to begin with, the strictly time-reversible case, the corresponding dissipation function (2.3) must vanish identically. Thus, if one allows for $\Lambda$ to be defined by the restriction to $\Lambda = \Lambda(\mathbf{x},t)$, then upon the combination of Eqs. (3.1.4) and (3.1.5), Eq. (3.1.2) follows immediately.

Finally, the proposal that the frequency $\omega$ is maintained in a steady-state via the constant throughput of thermal energy has to be cast into a re-formulation of what is understood as "total energy", i.e., of Eq. (1.1). For the time being, we do not need to specify what exactly this thermal energy is, although it is likely related to the vacuum, and one can think of some good candidates here. (Think, for example, of the vacuum's zero-point energy, or of the recently established evidence that the universe is permeated by some form of "dark energy".) All we need to specify in the beginning is that a quantum system's energy consists of the "total energy" of the "system of interest" (i.e., the particle with frequency $\omega$), and of some term representing energy throughput related to the surrounding vacuum, i.e., effectively some function $F$ of the heat flow $\Delta Q$:

$$E_{\text{tot}}(\mathbf{x},t) = E(\omega,\mathbf{x},t) + F[\Delta Q(\mathbf{x},t)]. \quad (3.1.6)$$

The first term is assumed to be given by Eq. (1.1), and the second term, being equivalent to some kinetic energy, can be recast with the aid of a fluctuating momentum term, $\delta \mathbf{p}$, of the particle with momentum $\mathbf{p}$. Thus, the total energy is given by (**Assumption 3**):

$$E_{tot} = \hbar \omega + \frac{(\delta p)^2}{2m}. \tag{3.1.7}$$

That is all we need: Eqs. (3.1.1), (3.1.2) and (3.1.7) suffice to derive the exact Schrödinger equation from (modern) classical physics. This shall be shown now.

### 3.2 Derivation of the exact Schrödinger equation from classical physics

We consider the standard Hamilton-Jacobi formulation of classical mechanics, with a "total internal energy" of the system of interest generally given by

$$\hbar \omega = \sum_i p_i^2 / 2m + V(x), \tag{3.2.1}$$

where $V$ is some potential energy. In the following, we shall for simplicity restrict ourselves to the one-particle case $(i=1)$, as an extension to the many-particle case can easily be done.

We introduce the action function $S(\mathbf{x},t)$ such that the total energy of the whole system (i.e., our "system of interest" and the additional kinetic energy due to the assumed heat flow) is given by

$$E_{tot}(\mathbf{x},t) = -\frac{\partial S(\mathbf{x},t)}{\partial t}. \tag{3.2.2}$$

To start with, we consider as usual the momentum $\mathbf{p}$ of the particle as given by

$$\mathbf{p}(\mathbf{x},t) = m\mathbf{v}(\mathbf{x},t) = \nabla S(\mathbf{x},t), \tag{3.2.3}$$

noting, however, that this will not be the effective particle momentum yet, due to the additional momentum coming from the heat flow. With these preliminary definitions, we formulate the action integral in an $n-$ dimensional configuration space with the Lagrangian $L$ as

$$A = \int L\, d^n x dt = \int P(\mathbf{x},t) \left[ \frac{\partial S}{\partial t} + \frac{1}{2m} \nabla S \cdot \nabla S + \frac{1}{2m} \nabla(\delta S) \cdot \nabla(\delta S) + V \right] d^n x dt, \tag{3.2.4}$$

where we have introduced the momentum fluctuation of Eq. (3.1.7) as

$$\delta \mathbf{p} = \nabla(\delta S). \tag{3.2.5}$$

Our task is now to derive an adequate expression for $\delta \mathbf{p}$ from our central assumption, i.e., from an underlying nonequilibrium thermodynamics. To begin, we remember the distinction between "heat" as disordered internal energy on one hand, and mechanical work on the other: heat as disordered energy cannot be transformed into useful work by any means. According to Boltzmann, if a particle trajectory is changed by some supply of heat $\Delta Q$ to the system, this heat will be spent either for the increase of disordered internal energy, or as ordered work furnished by the system against some constraint mechanism [6]:

$$\Delta Q = \Delta E_{\text{internal}} + \Delta W_{\text{constraints}}. \tag{3.2.6}$$

With $\Delta W_{\text{constraints}}$ being the effect of a heat flow, Eq. (3.2.6) is a corollary of Eq. (2.7), where the work applied to the system effectively produces a heat flow. This is why $\Delta W$ has different signs in the two respective equations. However, we first want to concentrate on time-reversible scenarios where $\Delta W = 0$.

It is clear that for the limiting case of Hamiltonian flow, which is characterized by a vanishing phase space contraction (3.1.5), a time-reversible scenario is evoked where $\Delta W = 0$ for all times. However, one can also maintain time reversibility by choosing that only the time *average* vanishes, $\overline{\Delta W} = 0$, thus allowing for the system of interest to be a nonequilibrium steady-state one. So, in what follows we shall at first restrict ourselves to the case where *on average* no work is done, $\overline{\Delta W} = 0$, which is equal to the time-reversible scenario. In the consecutive Chapter, then, we shall consider the time-irreversible case, $\overline{\Delta W} \neq 0$.

If in Eq. (2.7), or Eq. (2.8), respectively, we therefore set $\Delta W = 0$ (which due to the specific form of these Equations *per se* already implies time averaging), the dissipation function $\overline{\Omega_t}$ vanishes identically, which in turn confirms time reversibility. However, as $\overline{\Omega_t} = 0$, we obtain with Eqs. (2.3), (2.5), and (3.1.2) the probability (density) ratio

$$\frac{P(\mathbf{x},t)}{P(\mathbf{x},0)} = e^{-\frac{\Delta Q}{kT}}. \tag{3.2.7}$$

This is equivalent to the form of the usual Maxwell-Boltzmann distribution for thermodynamical equilibrium, but this time it is the result of universal thermostatting in nonequilibrium thermodynamics under the restriction that *on average* the work vanishes identically.

Now, in order to proceed in our quest to obtain an expression for the momentum fluctuation (3.2.5) from our thermodynamical approach, we can again rely on a formula originally derived by Ludwig Boltzmann. As mentioned above, Boltzmann considered the change of a trajectory by the application of heat $\Delta Q$ to the system.

Considering a very slow transformation, i.e., as opposed to a sudden jump, Boltzmann derived a formula which is easily applied to the special case where the motion of the system of interest is oscillating with some period $\tau = 2\pi/\omega$. Boltzmann's formula relates the applied heat $\Delta Q$ to a change in the action function $S = \int (E_{kin} - V)\, dt$, i.e., $\delta S = \delta \int E_{kin} dt$, providing [6,7]

$$\Delta Q = 2\omega \delta S = 2\omega \left[ \delta S(t) - \delta S(0) \right]. \tag{3.2.8}$$

This is in perfect agreement with the standard relation for integrable conservative systems, which we do deal with as long as we restrict ourselves to considering properties of our "system of interest", providing an invariant action function $I = \int 2\delta E_{kin} dt = 2\delta S$. As originally proposed by Ehrenfest and reformulated in Goldstein [8],

$$dI = \frac{dE}{\omega}. \tag{3.2.9}$$

Identifying $dE$ with the heat flow $\delta Q$, and with $I = 2\delta S$ as just mentioned, Eq. (3.2.9) provides exactly the relation (3.2.8) again. (We shall return to Eq. (3.2.9) in Chapter 4, when we discuss the extension of our approach to non-conservative systems.)

Note that in Eq. (3.2.8) we already have obtained a connection between the heat flow $\Delta Q$ and our looked-for momentum fluctuation $\delta \mathbf{p}$, the latter being given by Eq. (3.2.5), $\delta \mathbf{p} = \nabla(\delta S)$. What remains to be identified with familiar expressions, is the term $kT$ in Eq. (3.2.7). It refers to the apparent temperature of the surroundings of our system of interest, with the latter having a total internal energy $\hbar\omega$.

Now, just as Eq. (3.2.7) was derived from very general, i.e., model-independent, features of nonequilibrium thermodynamics ("universal thermostatting"), we can now also give an alternative expression for the temperature of the thermostat from a very general observation. The latter is concerned with a universal property of harmonic oscillators: All sinusoidal oscillations have the simple property that the average kinetic energy is equal to half of the total energy. [9] Now, our system of interest has the total internal energy of $E = \hbar\omega$, and we deal with steady-state systems where the internal temperature on average matches the external one of the surrounding medium. We thus obtain with the requirement that the average kinetic energy of the thermostat, Eq. (2.2), must equal the average kinetic energy of the oscillator, that for each degree of freedom

$$\frac{kT}{2} = \frac{\hbar\omega}{2}. \quad (3.2.10)$$

Combining, therefore, Eqs. (3.2.7), (3.2.8), and (3.2.10), we obtain

$$P(\mathbf{x},t) = P(\mathbf{x},0) e^{-\frac{2}{\hbar}\left[\delta S(\mathbf{x},t) - \delta S(\mathbf{x},0)\right]}. \quad (3.2.11)$$

Thus we obtain from Eq. (3.2.11) our final expression for the momentum fluctuation $\delta \mathbf{p}$, derived exclusively from model-independent universal features of harmonic oscillators and nonequilibrium thermodynamical systems:

$$\delta\mathbf{p}(\mathbf{x},t) = \nabla\left(\delta S(\mathbf{x},t)\right) = -\frac{\hbar}{2}\frac{\nabla P(\mathbf{x},t)}{P(\mathbf{x},t)}. \quad (3.2.12)$$

This further provides the expression for the additional kinetic energy term in Eq. (3.2.4), i.e.,

$$\delta E_{\text{kin}} = \frac{1}{2m}\nabla(\delta S)\cdot\nabla(\delta S) = \frac{1}{2m}\left(\frac{\hbar}{2}\frac{\nabla P}{P}\right)^2. \quad (3.2.13)$$

As will be shown shortly, inserting Eq. (3.2.13) into the action integral (3.2.4) will ultimately provide the Schrödinger equation. (For an earlier version, see ref. [10], where also Heisenberg's uncertainty principle is derived from Eq. (3.2.12).)

Before doing so, the following remark may be helpful. There is an alternative way to derive the final action integral by referring in Eq. (3.2.4) to a generalized average momentum $\overline{\mathbf{p}} = \overline{\nabla(S+\delta S)} = \nabla S + \overline{\delta \mathbf{p}}$ instead of the two kinetic energy terms. Then, instead of Eq. (3.2.4), there would only remain one term for the kinetic energy, given by $\frac{1}{2m}\overline{\mathbf{p}} \cdot \overline{\mathbf{p}}$. However, as the average momentum fluctuations $\overline{\delta \mathbf{p}}$ must be linearly uncorrelated with the average momentum $\overline{\nabla S}$, such that the (averaged) vector product is unbiased [11], one has

$$\int P \overline{(\nabla S \cdot \delta \mathbf{p})} \, d^n x = 0, \qquad (3.2.14)$$

such that the terms with mixed momentum components vanish identically and the action integral again is given by Eq. (3.2.4). In fact, the requirement (3.2.14) is immediately obtained also from our requirement that the dissipation function, or the average work, respectively, vanishes identically. For, if we identify in Eq. (2.8) the flux $\mathbf{J}$ as the probability density current, i.e.,

$$\mathbf{J} = P\overline{\mathbf{v}} = P\frac{\overline{\nabla S}}{m}, \qquad (3.2.15)$$

and if we characterize the external force $\mathbf{F}_e$ by the change in momentum $\overline{\delta \mathbf{p}}$, i.e.,

$$\mathbf{F}_e = m\frac{\overline{\delta \mathbf{v}}}{\delta t} = \frac{\overline{\delta \mathbf{p}}}{\delta t}, \qquad (3.2.16)$$

the average work, assuming ergodicity, is given by

$$\frac{1}{t}\overline{\Delta W} = -\frac{1}{\overline{V}} \int d^n x P \frac{1}{m} \overline{(\nabla S \cdot \delta \mathbf{p})} = 0, \qquad (3.2.17)$$

which thus confirms Eq. (3.2.14), and, ultimately, the action integral (3.2.4). (We also note here that *a posteriori*, with the quantum physical equations already at our disposal, one can provide an additional, compulsory argument that necessarily confirms Eq. (3.2.17), as will be shown later.) This concludes the remark.

Returning to our main line of reasoning, we now turn to the derivation of the Schrödinger equation. We begin by recalling the identity (3.1.1), i.e. $P = R^2$, of the probability density with the intensity of waves of amplitude $R$. (Note: This holds for the time-reversible scenario, which we deal with here. In general, this identity does not necessarily hold for nonequilibrium situations. [12]) Thus, the action integral we have arrived at now reads

$$A = \int P \left[ \frac{\partial S}{\partial t} + \frac{\overline{p_{tot}}^2}{2m} + V \right] d^n x dt, \qquad (3.2.18)$$

where

$$\overline{p_{tot}} = \overline{\hbar \mathbf{k}_{tot}} =: \overline{\hbar \mathbf{k} + \hbar \mathbf{k}_u} = \overline{\nabla(S + \delta S)} = \nabla S - \hbar \overline{\frac{\nabla R}{R}}. \qquad (3.2.19)$$

Now we introduce the "Madelung transformation" (with the star denoting complex conjugation),

$$\psi^{(*)} = R e^{(-)\frac{i}{\hbar}S}. \qquad (3.2.20)$$

Thus one has

$$\frac{\nabla \psi}{\psi} = \frac{\nabla R}{R} + \frac{i}{\hbar} \nabla S, \text{ and } \left|\frac{\nabla \psi}{\psi}\right|^2 = \left(\frac{\nabla R}{R}\right)^2 + \left(\frac{\nabla S}{\hbar}\right)^2, \qquad (3.2.21)$$

and one obtains a *transformation rule between the formulations of modern classical physics and orthodox quantum theory*: the square of the average total momentum is given by

$$\overline{p_{tot}}^2 = \hbar^2 \left[ \left(\frac{\nabla R}{R}\right)^2 + \left(\frac{\nabla S}{\hbar}\right)^2 \right] = \hbar^2 \left|\frac{\nabla \psi}{\psi}\right|^2. \qquad (3.2.22)$$

With $P = R^2 = |\psi|^2$ from equation (3.2.20) one can rewrite (3.2.18) as

$$A = \int L dt = \int d^n x dt \left[ |\psi|^2 \left(\frac{\partial S}{\partial t} + V\right) + \frac{\hbar^2}{2m} |\nabla \psi|^2 \right]. \qquad (3.2.23)$$

Further, with the identity

$$|\psi|^2 \frac{\partial S}{\partial t} = -\frac{i\hbar}{2} \left(\psi^* \dot{\psi} - \dot{\psi}^* \psi\right)$$

one finally obtains the well-known Lagrange density

$$L = -\frac{i\hbar}{2} \left(\psi^* \dot{\psi} - \dot{\psi}^* \psi\right) + \frac{\hbar^2}{2m} \nabla \psi \cdot \nabla \psi^* + V \psi^* \psi. \qquad (3.2.24)$$

As given by the standard procedures of classical physics, this Lagrangian density provides (via the Euler-Lagrange equations) the Schrödinger equation

$$i\hbar \frac{\partial \psi}{\partial t} = \left(-\frac{\hbar^2}{2m} \nabla^2 + V\right) \psi. \qquad (3.2.25)$$

Without knowledge of the course of physics during the twentieth century, one might wonder why one had to introduce the Madelung transformation (3.2.20) in the first place. For, remaining within the language of classical physics would have also provided a correct and useful answer: Rewriting the action integral (3.2.18), or, respectively, (3.2.4) with the specification of Eq. (3.2.12) or Eq. (3.2.13), i.e.,

$$A = \int P(\mathbf{x},t) \left[\frac{\partial S}{\partial t} + \frac{(\nabla S)^2}{2m} + \frac{\hbar^2}{8m}\left(\frac{\nabla P}{P}\right)^2 + V\right] d^n x dt, \qquad (3.2.26)$$

one obtains upon fixed end-point variation in $S$ the usual continuity equation (3.1.3), and, more importantly, upon variation in $P$, a modified Hamilton-Jacobi equation,

$$\frac{\partial S}{\partial t} + \frac{(\nabla S)^2}{2m} + V + U = 0, \qquad (3.2.27)$$

where $U$ is known as the "quantum potential"

$$U = \frac{\hbar^2}{4m}\left[\frac{1}{2}\left(\frac{\nabla P}{P}\right)^2 - \frac{\nabla^2 P}{P}\right] = -\frac{\hbar^2}{2m}\frac{\nabla^2 R}{R}. \qquad (3.2.28)$$

Eqs. (3.1.3) and (3.2.27) form a set of coupled differential equations and thus provide the basis for the de Broglie-Bohm interpretation [13, 14], which can give a causal account of quantum motion. Still, as is well known, these two differential equations can, with the aid of the Madelung transformation (3.2.20) be condensed into a single differential equation, i.e., the Schrödinger equation (3.2.25), from which, historically, they were originally derived. So, the answer to the question, "why the Madelung transformation?", lies in the compactness of the single equation, and, most importantly, in its linearity: the Madelung transformation is a means to linearize an otherwise highly nonlinear set of coupled differential equations. Thus, the Schrödinger equation has the distinct advantage of an easy handling of the mathematics, although the disadvantage is given by the fact that $\psi(\mathbf{x},t)$ has no direct physical meaning, as opposed to all the quantities given in the Equations (3.1.3) and (3.2.27).

What is new in the present paper, though, is the result that *all* these latter quantities are, in fact, derived from "modern classical" ones, i.e., also the term $U$. For, as we have seen, the new input (i.e., as opposed to ordinary classical mechanics without any embedding of systems of interest in nonequilibrium processes) is an additional term for the kinetic energy, Eq. (3.2.13), $\delta E_{\text{kin}} = \frac{1}{2m}\left(\frac{\hbar}{2}\frac{\nabla P}{P}\right)^2$, which in the variational problem as shown above provides the quantum potential term

$$U = \frac{\hbar^2}{4m}\left[\frac{1}{2}\left(\frac{\nabla P}{P}\right)^2 - \frac{\nabla^2 P}{P}\right] = \frac{m\mathbf{u}\cdot\mathbf{u}}{2} - \frac{\hbar}{2}(\nabla\cdot\mathbf{u}) = \frac{\hbar^2}{2m}(\mathbf{k}_\mathbf{u}\cdot\mathbf{k}_\mathbf{u} - \nabla\cdot\mathbf{k}_\mathbf{u}), \qquad (3.2.29)$$

where

$$\mathbf{u} := \frac{\delta \mathbf{p}}{m} = -\frac{\hbar}{2m}\frac{\nabla P}{P} \quad \text{and} \quad \mathbf{k}_{\mathbf{u}} = -\frac{1}{2}\frac{\nabla P}{P} = -\frac{\nabla R}{R}. \tag{3.2.30}$$

Thus, we see that the expression "quantum potential" is rather misleading, since the term derives from a kinetic energy, and does indeed exactly represent a kinetic energy term, $\frac{mu^2}{2}$, in the case that $\nabla \cdot \mathbf{u} = 0$. Still, we shall accept and retain the name in the following, because it is so often used and well-known in the literature. The reader is referred to excellent reviews (e.g., [13, 14]) for discussions on the properties of $U$, of which we here want to mention the one very particular feature, namely, that it does not necessarily fall off with the distance, i.e., it is made "responsible" for the nonlocal effects of quantum theory. This is so despite another remarkable property, which actually is founded in very basic information theoretic principles [15], i.e., that its average spatial gradient vanishes identically:

$$\int P \, \nabla U \, d^3x = 0. \tag{3.2.31}$$

Moreover, differentiation of Eq. (3.2.27) provides the equations of quantum motion [13,14]:

$$m\frac{d\mathbf{v}}{dt} = -\nabla(V + U). \tag{3.2.32}$$

This confronts us with an intriguing observation: apart from the gradient of the classical potential, which just results in a classical force term affecting the momentum $\nabla S$ of the "internal" part of our system of interest, the (nonlocal) quantum potential is exactly the reason for an acceleration of the particle due to a "contextual" dynamics from outside the immediate (classical) system of interest. If we thus put in Equ. (3.2.16)

$$\mathbf{F}_e = \frac{\delta \mathbf{p}}{\delta t} = -\nabla U \tag{3.2.33}$$

and insert this into the defining equation of the work applied to our system, Eq. (2.8), we obtain (with $\mathbf{J} = P\mathbf{v}$ as before)

$$\Delta W = \int_0^t ds \mathbf{v} \widetilde{\nabla} P \nabla U = \int_{\mathbf{x}_0}^{\mathbf{x}_1} d\mathbf{x} \int d^3 x P \nabla U = 0. \tag{3.2.34}$$

This confirms that time reversibility is equivalent to both a vanishing average gradient of the quantum potential (due to Eq. (3.2.31)) and a vanishing average work applied to the system of interest, i.e., the particle of total (internal) energy $E = \hbar \omega$. Moreover, as the average external force

$$\mathbf{F}_e := \langle -\nabla U \rangle = -\int P \nabla U d^3 x = 0 \tag{3.2.35}$$

for time-reversible systems *in general*, Eq. (3.2.34) is another justification, this time *a posteriori*, of the average orthogonality of the vectors $\overline{\mathbf{p}}$ and $\overline{\delta \mathbf{p}}$ as given in Eq. (3.2.14), or in Eq. (3.2.17), respectively.

Finally, it should be noted that although the present derivation deals only with spinless particles, it is not only its historical priority which demands that the genuine Schrödinger equation be considered as the most essential equation of quantum theory. Just as a possible extension to relativistic cases, the extensions to include spinning particles must be on the agenda as "next steps", which can only be made, in the context presented here, after the foundations of the Schrödinger equation have become clear.

### 4. Extension to integrable nonconservative systems and the Vacuum Fluctuation Theorem

Now we want to extend our scheme to include integrable nonconservative systems. This means that the average work applied to the system of interest will not vanish, $\overline{\Delta W} \neq 0$, and also the average fluctuating quantum force $\overline{\mathbf{F}_e} = \langle -\nabla(\delta U) \rangle \neq 0$. Thus, assuming still the validity of the "internal" equilibrium implied by Eq. (3.1.2), we obtain from Eqs. (2.7) and (2.8) that

$$\overline{\Omega_t} t = \frac{\overline{\Delta W}}{kT} = -\ln \frac{P(\mathbf{x},t)}{P(\mathbf{x},0)} - \frac{\Delta Q}{kT} = -\frac{1}{kT} \int dt P \widetilde{V} \, \overline{\mathbf{v}} m \frac{\delta \mathbf{v}}{\delta t}, \qquad (4.1)$$

where the expression on the r.h.s. equals, analogously to Eq. (3.2.34),

$$\overline{\Omega_t} t = \frac{\overline{\Delta W}}{kT} = -\frac{1}{kT} \int d\mathbf{x} \langle -\nabla(\delta U) \rangle = \frac{1}{kT} \overline{\delta U}. \qquad (4.2)$$

With Eq. (4.1) we obtain the generalization of Eq. (3.2.7) as

$$P(\mathbf{x},t) = P(\mathbf{x},0) e^{-\frac{1}{kT}(\Delta Q + \overline{\Delta W})} = P(\mathbf{x},0) e^{-\frac{1}{kT}(\Delta Q + \overline{\delta U})}. \qquad (4.3)$$

As $\Delta Q$ refers to the heat applied to our system of interest and is given by Eq. (3.2.8), and as $\overline{\delta U}$ refers to an additional non-vanishing external energy, we also obtain, with $\frac{1}{\omega} = \frac{\tau}{2\pi} = \delta t$, the generalization

$$\delta \mathbf{p}_{\text{tot}} = -\frac{\hbar}{2} \frac{\nabla P}{P} - \frac{\hbar}{2} \frac{\nabla(\delta P)}{\delta P} = \nabla \left[ \delta S + \frac{1}{2} \overline{\delta U} \delta t \right] =: \nabla(\delta S + \Delta S_{\text{ext}}), \qquad (4.4)$$

where the last term on the r.h.s. refers to a change in the "external" action due to a non-vanishing average fluctuation of the quantum potential. In terms of momenta, this means that an additional, external momentum $\delta \mathbf{p}_{\text{ext}} = \nabla(\Delta S_{\text{ext}})$ must be added in the balance (3.2.30) to provide the new total momentum fluctuation

$$\delta \mathbf{p}_{\text{tot}} = m\mathbf{u} + \delta \mathbf{p}_{\text{ext}}. \qquad (4.5)$$

We shall return to Eq. (4.5) below, when we discuss implications of the vacuum fluctuation theorem. Here we just note that, alternatively, $\delta \mathbf{p}_{\text{tot}}$ can also be written as

$$\delta \mathbf{p}_{\text{tot}} = -\frac{\hbar}{2} \frac{\nabla(P \delta P)}{P \delta P} = -\frac{\hbar}{2} \left[ 3 \frac{\nabla R}{R} + \frac{\nabla(\delta R)}{\delta R} \right]. \qquad (4.6)$$

As is well known, Hamilton's principle applies for both conservative and nonconservative systems, i.e.,

$$\delta S + E(t)\delta t = 0, \qquad (4.7)$$

where fixed end-points are assumed and the Lagrange multiplier $E(t)$ is the true value of the energy at time $t$ (i.e., after having the particle path starting at time $t=0$). Gray *et al.*, in an extensive survey of variational principles [17], provide a so-called "unconstrained Maupertius principle" (UMP) for nonconservative systems, which relates the variations of a mean energy $\overline{E}$, of action $S$, and of the travel time $t$, such that the Lagrange multipliers are the true travel time, and the difference between energy and mean energy of the true trajectory at time $t$, the latter being

$$E(t) - \overline{E}(t) = \frac{1}{t}\int dt \left[ t \frac{\partial H}{\partial t} \right] =: \left\langle t \frac{\partial H}{\partial t} \right\rangle. \qquad (4.8)$$

Now, let us turn to our "system of interest", i.e., our oscillating particle with period $t$, and with the action $I$ as an adiabatic invariant obeying Eq. (3.2.9),

$$dI = \frac{dE}{\omega} = dE \frac{t}{2\pi}. \qquad (4.9)$$

For such periodic systems, both the energy $E$ and the period $t$ are functions of the action $I$. If one now compares two actual trajectories with action $I$ and $I+dI$ as two particular ones, the above mentioned UMP can be written as [17]

$$\frac{d\overline{E}}{dI} - \frac{2\pi}{t} = \left\langle t \frac{\partial E}{\partial t} \right\rangle \frac{1}{t}\frac{dt}{dI}. \qquad (4.10)$$

In terms of the frequency $\omega$, Eq. (4.10) reads as

$$dI = \frac{d\overline{E}}{\omega} + \left\langle t \frac{\partial E}{\partial t} \right\rangle \frac{1}{\omega^2} d\omega, \qquad (4.11)$$

which reduces to Eq. (4.9) for conservative systems. Remembering from Chapter 3 that for our periodic system $I = \int 2\delta E_{\text{kin}} dt =: 2\delta S$, one can also write with $\left\langle t \frac{\partial E}{\partial t} \right\rangle := \overline{\delta_t E}$

$$\frac{d\overline{E}}{\omega} = 2\delta S - \frac{\overline{\delta_t E}}{\omega} \frac{d\omega}{\omega}. \tag{4.12}$$

Whereas we therefore have for conservative systems with Eq. (3.2.8) that $\Delta W = dE - 2\omega \delta S = 0$, we now have for nonconservative systems

$$\Delta W = d\overline{E} - 2\omega \delta S = -\overline{\delta_t E} \frac{d\omega}{\omega}. \tag{4.13}$$

To illustrate the meaning of Eq. (4.13), an example of a nonconservative system has been studied to show that the results still compare with those of the usual quantum mechanics. It is given in an extended version of the present paper and will be published elsewhere [18].

In the preceding chapters, we have seen that nonequilibrium thermodynamics is a very useful field that can be employed for a deeper understanding of quantum theory. Now, we do of course not know much about the peculiarities of the hypothesised sub-quantum medium. There exists, for example, the possibility that the application of the formalism regarding the dissipation function was, in fact, correct, but the broader theory regarding the fluctuation theorem (FT) was not. This (rather minute) possibility notwithstanding, and in view of the actual successful application of nonequilibrium thermodynamics so far, one can consider it encouraging enough to also probe the more encompassing statements of the FT and try to apply them on the sub-quantum level.

Referring, then, to Eq. (2.4), which is a formulation of the TFT for steady-state systems, we can re-formulate said equations in terms of the variables employed in the (sub-)quantum domain. From Eq. (4.2) we get with Eq. (3.2.10) that

$$\overline{\Omega_t} t = \frac{\overline{\Delta W}}{kT} = \frac{\overline{\delta U}}{\hbar \omega}. \qquad (4.14)$$

Moreover, we note that generally, with $\overline{\delta(\nabla^2 R)} \equiv 0$,

$$\overline{\delta U} = -\overline{\left(\frac{\delta R}{R}\right) U}. \qquad (4.15)$$

Then, we can formulate a TFT which is assumed to hold for the vacuum (thermo-)dynamics of the (sub-)quantum domain, and which we call the *Vacuum Fluctuation Theorem* (VFT):

$$\frac{p\left(\overline{\Omega_t} = \frac{1}{t}\frac{\overline{\delta U}}{\hbar \omega} = A\right)}{p\left(\overline{\Omega_t} = \frac{1}{t}\frac{\overline{\delta U}}{\hbar \omega} = -A\right)} = e^{At} = e^{\overline{\delta U}/\hbar \omega}. \qquad (4.16)$$

With Eq. (4.15), we write

$$At = \frac{\overline{\delta U}}{\hbar \omega} = -\overline{\left(\frac{\delta R}{R}\right) \frac{U}{\hbar \omega}} =: -\overline{\left(\frac{\delta R}{R}\right) \tilde{A}}, \qquad (4.17)$$

and we obtain (with an obvious notational shorthand)

$$\frac{p(A)}{p(-A)} = e^{-\overline{(\delta \ln R) \tilde{A}}} = e^{-\tilde{A}} \overline{\left(\frac{R + \delta R}{R}\right)}. \qquad (4.18)$$

Note that, for example, in the problem of the "particle in a box", $\tilde{A} = \frac{\overline{U}}{\hbar \omega} = 1$, such that Eq. (4.18) is no more characterized by an exponential relationship between $p(A)$ and $p(-A)$, respectively, but rather that fluctuations $\delta R$ can have relatively high probabilities both for the $A$ and the $-A$ cases, respectively. Generally, we have from Eq. (4.18) upon re-insertion of (4.15) that

$$\frac{p(A)}{p(-A)} = e^{-\frac{\overline{U}}{\hbar\omega}}\left(1 - \frac{\overline{\delta U}}{\overline{U}}\right). \tag{4.19}$$

A more detailed discussion of the implications of Eq. (4.19) will be given in a forthcoming paper. For now it shall suffice to have a look at the following consequence. As with Eq. (4.4) we have that

$$\delta \mathbf{p}_{ext} := -\frac{\hbar}{2}\frac{\nabla(\delta P)}{\delta P} = \frac{1}{2}\nabla\left(\overline{\delta U}\delta t\right) = \frac{\hbar}{2}\nabla(At), \tag{4.20}$$

we obtain with Eq. (4.16) that

$$\delta \mathbf{p}_{ext} = \frac{\hbar}{2}\nabla\left\{\ln\frac{p(A)}{p(-A)}\right\} = \frac{\hbar}{2}\left\{\frac{\nabla p(A)}{p(A)} - \frac{\nabla p(-A)}{p(-A)}\right\}. \tag{4.21}$$

Thus, the total momentum fluctuation due to Eq. (4.5) is

$$\delta \mathbf{p}_{tot} = -\frac{\hbar}{2}\nabla \ln\{P + p(-A) - p(A)\}. \tag{4.22}$$

The first term on the r.h.s. of Eq. (4.22) refers to the usual momentum fluctuation $\delta \mathbf{p} = m\mathbf{u}$ (i.e., which leads to the quantum potential term in the modified Hamilton-Jacobi equation). However, the second and third terms refer to fluctuations of the overall system in which our "system of interest" is embedded. Here, it is crucial that these fluctuations, according to the VFT, can in principle be arbitrarily large! We also see that even for the cases that $p(A)$ or $p(-A)$ are very small by themselves, the relative gradients $\nabla p/p$ can provide significant contributions to $\delta \mathbf{p}_{ext}$. If we consider, for example, the time-dependent term $\overline{\delta U} \neq 0$ in the case of a delayed-choice experiment, which is a prototype of an experiment that can be characterized by "moving walls" of an experimental configuration [19], there may emerge significant contributions to momentum fluctuations, $-\frac{\hbar}{2}\overline{\left(\frac{\nabla(\delta R)}{\delta R}\right)}$, even as a result of minimal changes of amplitudes $\overline{\delta R}$ over arbitrary distances within the confines of the "box", i.e., the experimental setup between source and detectors. This, then, is a strong indication that the vacuum alone can serve as a resource for entanglement. The VFT can thus possibly provide a framework for the deeper understanding of how, or why, entanglement can come about. Moreover, possible experimental tests of the VFT are conceivable which may reach beyond the scope of present-day quantum theory.

**Conclusions and outlook**

It was shown that by merging nonequilibrium thermodynamics with only a few basics of classical wave mechanics, the exact Schrödinger equation can be derived, and a general "Vacuum Fluctuation Theorem" (VFT) regarding vacuum fluctuations responsible for quantum effects can be formalized. Note that in the course of this derivation, apart from the **Assumptions 1 – 3**, no parameter adjustments were made, or any other form of "guessing" of constants, approximations, etc. As, for example, in Nelson's derivation of the Schrödinger equation, the "diffusion constant" $D := \frac{\hbar}{2m}$ is put in "by hand" [16], we claim that here no such extra assumptions are necessary. This leads us to the claim that the present work exhibits the "fastest" way to derive the exact Schrödinger equation from modern classical physics.

Specifically, we have identified a dissipative force field $\mathbf{F_e}$ as being due to the action of the "quantum potential", $\mathbf{F_e} = \langle -\nabla U \rangle$, which vanishes identically for conservative systems, but $\mathbf{F_e} = \langle -\nabla \overline{\delta U} \rangle \neq 0$ for nonconservative systems. The "quantum potential" is given by $U = \frac{m\mathbf{u} \cdot \mathbf{u}}{2} - \frac{\hbar}{2}(\nabla \cdot \mathbf{u})$, where $\mathbf{u}$ can be written as either $\mathbf{u} = -\frac{\hbar}{2m}\frac{\nabla P}{P}$, or, equivalently, via Eqs. (3.2.8) and (3.2.11), as $\mathbf{u} = \frac{1}{2\omega m}\nabla Q$, which thus clearly exhibits its dependence on the spatial behaviour of the heat flow $\delta Q$.

Throughout this paper, though, we have refrained from speculations about the nature of the latter, which ultimately may well be cosmological, or of cosmological significance, respectively. In perhaps the simplest scenario, $Q$ could just refer to the heat that is around everywhere in what are called the vacuum's zero-point fluctuations, whose energy content is given exactly by the amount of Eq. (3.2.10). Steady-state systems, then, would on average absorb some amount of heat and release it again, thus maintaining a constant temperature of their environment.

Finally, note that in the present paper no attempt is made to explain the appearance of Planck's constant. However, there already do exist some highly interesting approaches in the literature which strongly suggest that also $\hbar$ can be understood within the domain of a properly expanded, but basically classical physics. For example, Timothy Boyer has shown in a thermodynamic analysis of the harmonic oscillator that the Planck spectrum with zero-point radiation corresponds to the function satisfying the Wien displacement result which provides the smoothest possible interpolation between energy equipartition at low frequency and zero-point energy at high frequency. Equipartition theorems are also at the focus of Stephen Adler's theory, which, in fact, is the most elaborate attempt yet in the literature to explain quantum theory as emergent from an underlying classical theory. [21, 22] The latter is extended to non-commuting matrix variables, with cyclic permutation inside a trace as basic calculational tool. Quantum theory is shown to emerge as the statistical mechanics of this classical theory, with $\hbar$ and the canonical (anti-) commutation relations derived from it. It may well turn out that Adler's theory, focusing on the more formal features, has a direct correspondence to a more physical approach employing nonequilibrium thermodynamics, like the one that is attempted here.

**References:**

[1]     A. Einstein, „Über einen die Erzeugung und Verwandlung des Lichts betreffenden heuristischen Gesichtspunkt", *Ann. d. Phys.* **17** (1905) 132-148.

[2]     G. Gallavotti and E. G. D. Cohen, "Dynamical Ensembles in Stationary States", *Journal of Statistical Physics* **80** (1995) 931-970.

[3]     D. J. Evans and D. J. Searles, "The Fluctuation Theorem", *Advances in Physics* **51** (2002) 1529–1585.

[4]     S. R. Williams *et al.,* "Independence of the transient fluctuation theorem to thermostatting details", *Phys. Rev.* E **70,** 066113 (2004).

[5]     D. M. Carberry *et al.*, "Fluctuations and Irreversibility: An Experimental Demonstration of a Second Law-like Theorem Using a Colloidal Particle Held in an Optical Trap", *Phys. Rev. Lett.* **92**, 140601 (2004).

[6]     L. Boltzmann, „Über die mechanische Bedeutung des zweiten Hauptsatzes der Wärmetheorie", *Wien. Ber.* **53** (1866) 195–220.

[7]     L. Brillouin, *Tensors in Mechanics and Elasticity*, Academic Press, New York, 1964.

[8]     H. Goldstein *et al., Classical Mechanics*, 3$^{rd}$ edn., Addison Wesley, New York, 2002.

[9]     See, for example, R. P. Feynman *et al., The Feynman Lectures of Physics*, Addison Wesley, Reading, 1965.

[10] G. Grössing, "From classical Hamiltonian flow to quantum theory: derivation of the Schrödinger equation", *Found. Phys. Lett.* **17** (2004) 343–362. See also arXiv:quant-ph/0311109.

[11] M. J. W. Hall and M. Reginatto, "Schrödinger equation from an exact uncertainty principle", *J. Phys.* A **35** (2002) 3289-3303; see also ref. [10].

[12] A. Valentini, "Signal-Locality, Uncertainty, and the Subquantum H-Theorem. I and II", *Phys. Lett. A* **156** (1991) 263-268 and **158** (1991) 1-8.

[13] D. Bohm and B. J. Hiley, *The Undivided Universe*, Routledge, London, 1993.

[14] P. R. Holland, *The Quantum Theory of Motion*, Cambridge University Press, Cambridge, 1993.

[15] P. Garbaczewski, "Information dynamics and origins of uncertainty", arXiv: cond-mat/0703147.

[16] E. Nelson, "Derivation of the Schrödinger Equation from Newtonian Mechanics", *Phys. Rev.* **150** (1966) 1079. See also L. Fritsche and M. Haugk, "A New Look at the Derivation of the Schrödinger Equation from Newtonian Mechanics", *Ann. Phys. (Leipzig)* **12** (2003) 371-402.

[17] C. G. Gray *et al.*, "Progress in classical and quantum variational principles", *Rep. Prog. Phys.* **67** (2004) 159-208.

[18] G. Grössing, http://arxiv.org/abs/0711.4954 .

[19] G. Grössing, *Quantum Cybernetics*, Springer, New York, 2000.

[20] T. H. Boyer, "Thermodynamics of the Harmonic Oscillator: Wien's Displacement Law and the Planck Spectrum", *Am. J. Phys.* **71** (2003) 866-870.

[21] S. L. Adler and A. C. Millard, "Generalized Quantum Dynamics as Pre-Quantum Mechanics", *Nucl. Phys. B* **473**, 1 (1996) 199-244.

[22] S. L. Adler, *Quantum Theory as an Emergent Phenomenon*, Cambridge University Press, Cambridge, 2004.